\documentclass{article}
\usepackage{spconf,amsmath,graphicx,multirow,url}

\graphicspath{{figures/}}
\title{Phase recovery in NMF for audio source separation: an insightful benchmark}

\name{Paul Magron \qquad Roland Badeau \qquad Bertrand David \thanks{This work is partly supported by the French National Research Agency (ANR) as a part of the EDISON 3D project (ANR-13-CORD-0008-02).}}

\address{Institut Mines-T\'{e}l\'{e}com, T\'{e}l\'{e}com ParisTech, CNRS LTCI, Paris, France \\ \texttt{<firstname>.<lastname>@telecom-paristech.fr}}

\begin{document}
\ninept
\maketitle

\begin{abstract}
Nonnegative Matrix Factorization (NMF) is a powerful tool for decomposing mixtures of audio signals in the Time-Frequency (TF) domain. In applications such as source separation, the phase recovery for each extracted component is a major issue since it often leads to audible artifacts. In this paper, we present a methodology for evaluating various NMF-based source separation techniques involving phase reconstruction. For each model considered, a comparison between two approaches (blind separation without prior information and oracle separation with supervised model learning) is performed, in order to inquire about the room for improvement for the estimation methods. Experimental results show that the High Resolution NMF (HRNMF) model is particularly promising, because it is able to take phases and correlations over time into account with a great expressive power.

\end{abstract}


\section{Introduction}

The problem of separating polyphonic music mixtures into isolated sources has become very popular in the last 15 years. The family of techniques based on nonnegative factorizations, often applied to spectrogram-like representations, has proved to provide a successful and promising framework for this task~\cite{Smaragdis2003}.



NMF, originally introduced as a rank-reduction method \cite{Lee1999}, approximates a nonnegative data matrix $V$ as a product of two low-rank nonnegative matrices $W$ and $H$. In audio signal processing, $V$  is often chosen as the magnitude or power spectrogram of the signal, whose factorization is interpretable intuitively: $W$ is a dictionary of spectral templates and $H$ is a temporal activation matrix. Usual alternative versions constrain NMF to enforce properties such as sparsity \cite{Smaragdis2003}, smoothness or harmonicity \cite{Bertin2010, Rigaud2013}.

However, when it comes to resynthesize the separated time signals, the recovery of the phase of the corresponding Short-Time Fourier Transform (STFT) is necessary. Even if common practice consists in applying Wiener-like filtering (\emph{e.g} soft masking of the complex-valued STFT of the original mixture), phase recovery is still an open issue, for this kind of filtering does not enforce phase \emph{consistency}. That is, the obtained complex-valued matrix is not the STFT of a time signal.
It is worth noting here that  consistency can  also refer to specific properties of the instantaneous phase of a sinusoidal component \cite{Laroche1999}, but we will hereafter  employ \emph{consistency} in the first usage only.

Several extensions to NMF have been introduced, which include a phase model \cite{Fevotte2009,Kameoka2009,Ozerov2010}, but do not refer to phase consistency.
%
%
Wiener-like filtering is used for instance in \cite{Fevotte2009}. The separated components are then derived by inverting a TF representation whose phase is that of the STFT of the mixture.  This technique ensures phase consistency as long as only one source is active within each TF bin. In order to handle the case of overlapping sources, iterative methods \cite{Griffin1984, LeRoux2008} minimize the inconsistency of the reconstructed TF representation. On the other hand, some NMF-inspired models combine phase modeling and spectrogram factorization. The complex NMF model introduced in \cite{Kameoka2009} was later improved by means of consistency constraints \cite{LeRoux2009a}. More recently, High Resolution NMF (HRNMF) has been introduced in \cite{Badeau2011}. It models a TF mixture as a sum of autoregressive components in the TF domain, thus dealing explicitly with a phase model which takes time dependencies from one TF bin to another into account.

All the above-mentioned models are suitable for blind source separation of audio signals since they factorize the spectrogram, reconstruct the phase and enforce its consistency. In this paper, we propose a methodology for assessing their potential and performance.
This methodology is based on a comparison between two approaches: blind separation without prior information and oracle separation with supervised model learning. This comparison is performed in order to inquire about the room for improvement for the estimation methods. Algorithms are evaluated with \textsc{BSS Eval} \cite{Vincent2006}, a set of objective criteria dedicated to measuring source separation quality.
Finally, the algorithms are tested on different data types. Since difficulties often arise when sources overlap in the TF domain, a particular emphasis has been put on the related tests.

The paper is organized as follows. Section \ref{sec:algo} presents the considered NMF-based algorithms. Section \ref{sec:meth} describes the methodology of this benchmark, through objectives and protocol. Section \ref{sec:exp} presents results and interpretations of the tests conducted on a variety of data, and Section \ref{sec:conclu} draws some concluding remarks.

\section{NMF-based source separation algorithms}
\label{sec:algo}

\subsection{NMF main principle}

The NMF problem is expressed as follows: given a matrix $V$ of dimensions $F \times T$ with nonnegative entries, find a factorization $V \approx WH$ where $W$ and $H$ are nonnegative matrices of dimensions $F \times K$ and $K \times T$. In order to reduce the dimension of data, $K$ is chosen such that $K(F+T)\ll FT $. In audio source separation, $V$ is generally the magnitude or the power spectrogram of a TF representation $X$ of a mixture signal (most of the time an STFT). One can interpret $W$ as a dictionary of spectral templates and $H$ as a matrix of temporal activations. If $W_k$ denotes the $k$-th column of $W$ and $H_k$ denotes the $k$-th line of $H$, then $V_k = W_kH_k$ is the magnitude or power spectrogram of the component indexed by $k$ and $\hat{V}=\sum_{k=1}^{K} V_k$. Note that this result expresses an additivity property of spectrograms, which only approximately holds when sources overlap in the TF domain. This factorization is generally obtained by minimizing a cost function $D(V,\hat{V})$. Popular choices for $D$ are the Euclidean distance, Kullback-Leibler divergence \cite{Lee1999} and Itakura-Saito divergence \cite{Fevotte2009}. Our benchmark uses multiplicative update rules (MUR) \cite{Lee2001}, in order to estimate a regular NMF with the Kullback-Leibler divergence (KLNMF).

\subsection{Phase reconstruction}
Estimating a complex TF representation $X_k$ of a separated source by applying Wiener filtering \cite{Fevotte2009} consists in computing:

\begin{equation}
X_{k} = \frac{W_{k}H_{k}}{\sum_{l=1}^{K} W_{l}H_{l}}X = \frac{V_{k}}{\hat{V}}X.
\label{eq:wiener}
\end{equation}
This method will be referred to as \textbf{NMF-Wiener}.

Alternatively, a regular NMF can be combined with a phase reconstruction algorithm based on the minimization of a cost function which penalizes inconsistency. The \textbf{Griffin-Lim} algorithm \cite{Griffin1984} is an iterative method described in Eq. \eqref{eq:GL} for recursively estimating the $k$-th component. For each iteration $i$:

\begin{equation}
X^i_{k} \longrightarrow Y^{i+1}_{k}=\mathcal{F}(X^i_{k}) \longrightarrow X^{i+1}_{k}=\frac{V_{k}}{|Y^{i+1}_{k}|} Y^{i+1}_{k}
\label{eq:GL}
\end{equation}
where $ \mathcal{F} = STFT \circ STFT^{-1}$. It has been shown to make the Euclidean distance between $V_k$ and $| Y^{i}_{k}|$ decrease over iterations. This method will be referred to as \textbf{NMF-GL}.

The \textbf{LeRoux} algorithm \cite{LeRoux2008} consists in explicitly calculating and minimizing the inconsistency defined as the Euclidean distance between $X$ and $\mathcal{F}(X)$. Iterative optimization techniques then lead to update rules for the phase of the reconstructed source in the TF domain. This method will be referred to as \textbf{NMF-LR}.

In \textbf{NMF-GL} and \textbf{NMF-LR}, the magnitude is constant over iterations. The user can force it to be equal to $V_k$, obtained from the NMF. However, experiments show that initializing \textbf{Griffin-Lim} and \textbf{LeRoux} algorithms with the magnitude of $X_{k}$ in Eq. \eqref{eq:wiener} provides better results.

\subsection{Complex NMF}

Complex NMF \cite{LeRoux2009a} consists in factorizing a magnitude spectrogram while reconstructing a phase field for each source. The mixture TF representation is modeled as follows: for each TF bin $(f,t)$,

\begin{equation}
X(f,t) = \sum_{k=1}^{K} X_k(f,t)  = \sum_{k=1}^{K} W_k(f) H_k(t) e^{j\phi_{k}(f,t)}.
\end{equation}

This method will be referred to as \textbf{CNMF}. An explicit phase consistency constraint \cite{LeRoux2009a} leads to a consistent TF representation. It will be referred to as \textbf{CNMF-LR}. The main advantage of this technique is to jointly estimate the magnitude and phase parameters, instead of deriving the phase from an imposed magnitude (as in \textbf{NMF-LR}).

\subsection{High Resolution NMF}

More recently, the HRNMF model has been introduced in \cite{Badeau2011}. It consists in modeling each frequency band of the TF representation by means of auto-regressive filtering. This technique naturally captures phase relationships and dependencies over time.

The mixture TF representation is modeled as follows:
\begin{equation}
X(f,t) = n(f,t) + \sum_{k=1}^{K} X_{k}(f,t) 
\end{equation}
where $n(f,t)$ is a white Gaussian noise. Each source $X_{k}(f,t)$ is obtained by autoregressive filtering of a non-stationary signal $b_{k}(f,t)$:

\begin{equation}
X_k(f,t) = b_k(f,t) + \sum_{p=1}^{P(k,f)} a_p(k,f)X_k(f,t-p)
\end{equation}
where $P(k,f)$ is the order of the autoregressive filter for source $k$ and frequency $f$, of coefficients $a_p(k,f)$. Finally, $b_{k}(f,t)$ follows a centered normal distribution of variance $V_k(f,t)$ such that $V_k = W_k H_k$, and all $b_{k}(f,t)$ are independent.

The model parameters can be estimated either by a regular EM algorithm, which is computationally costly, or by a variational Bayesian EM (VBEM) algorithm, allowing faster computation without significant quality loss. We conduct an experience to estimate the best HRNMF initialization and algorithm in Section \ref{sec:HRNMF_init}. Note that recently, HRNMF has been extended to multichannel signals and convolutive mixtures, and is now able to model correlations over frequency~\cite{Badeau2014}.

\section{Methodology}
\label{sec:meth}

\subsection{Objectives}

In order to assess audio source separation quality, we use \textsc{BSS Eval} \cite{Vincent2006}, a set of objective criteria dedicated to this purpose. From the original sources $x_k$ and the estimated sources $\hat{x}_k$, $k=1,...,K$, \textsc{BSS Eval} computes various energy ratios: the SIR (signal to interference ratio) that measures the rejection of interferences, the SAR (signal to artifact ratio) for the rejection of artifacts, and the SDR (signal to distortion ratio) for the global quality.

In order to evaluate the room for improvement for these techniques, we compare the results obtained with a blind approach and an oracle approach. The blind approach consists in estimating the models directly from the mixture without using any prior information about the isolated sources. The oracle approach consists in evaluating, for each technique, the best performance possible: the parameters are learned from each isolated source. A comparison between those two approaches informs us about the opportunities for further enhancement of these methods. 

Since phase recovery is a major issue in source separation, it is interesting to evaluate if the consistency constraint used in various methods (\textbf{NMF-GL}, \textbf{NMF-LR} and \textbf{CNMF-LR}) is related to audio quality.

Finally, we want to evaluate the expressive power of the models, that is to say their ability to represent a variety of signals observed in music analysis. We use both synthetic and real data, with and without TF overlap.

The \textsc{MATLAB} code for this study is available at \url{http://perso.telecom-paristech.fr/magron/}.

\subsection{Datasets and protocol}

\begin{figure}[ht]
\centering
\hspace{-0.3cm}
\includegraphics[scale=0.4]{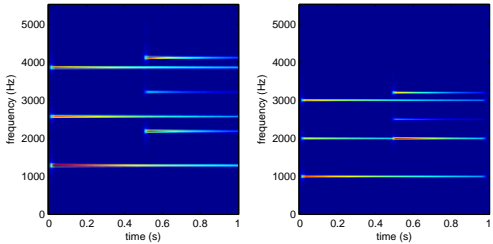}
\caption{Synthetic data spectrograms: without TF overlap (left) and with TF overlap (right)}
\label{fig:synth_data}
\end{figure}

We perform audio source separation on several datasets. Firstly, we synthesize $60$ mixtures of two harmonic signals ($K=2$) which consist of damped sinusoids whose amplitude, origin phase, frequency and damping coefficients are randomly-defined, and a $60$ dB additive white noise. The damping coefficient is the same for all harmonics. One set of $30$ mixtures does not include TF overlap while the other one does (see an example in Figure~\ref{fig:synth_data}).

The MAPS (MIDI Aligned Piano Sounds) dataset \cite{Emiya2010a} provides various data to design mixtures of real piano sounds. For the tests on real data, we consider $30$ mixtures of two piano notes, selected randomly in the MAPS database. We also enforce TF overlap in  some data.
Finally, we tested the benchmark on a $1.57$ second-long MIDI audio excerpt. It is composed of several occurrences of three bass notes and one guitar chord ($K=4$).

The data is sampled at $F_s = 11025$ Hz. It is important to note that HRNMF involves more diverse parameters than the regular NMF model. Indeed, correlations across time are taken into account by means of autoregressive filtering in each frequency sub-band of order $P(k,f)$. In our experiments, $P(k,f)$ was set to $1$ for all $(k,f)$. This means that the HRNMF model uses twice as many spectral parameters ($W$ and $a$) as regular NMF ($W$ only). In order to make a fair comparison, it is interesting to compare both models with the same total number of parameters. The STFT is thus calculated with a $512$ sample-long normalized Hann window with $75 \%$ overlap for testing \textbf{CNMF}, \textbf{CNMF-LR} and \textbf{HRNMF} models, and with a $1024$ sample-long window for testing \textbf{NMF-Wiener}, \textbf{NMF-GL} and \textbf{NMF-LR} models\footnote{Note that the total number of parameters involved in the CNMF model is higher than the dimension of the TF data itself, because all phase coefficients are free. However, even if comparing CNMF with NMF or HRNMF using the same total number of parameters is not possible, the results in Section~\ref{sec:exp} will show that CNMF is most often outperformed by the other models.}.

For both blind and oracle approaches, KLNMF and CNMF are estimated with 30 iterations of MUR algorithms, and phase reconstruction algorithms involve 50 iterations. HRNMF is initialized with a 30-iterations KLNMF and estimated with 30 iterations of the VBEM algorithm for the blind approach, and 10 iterations of the VBEM algorithm for each source learning (oracle approach). We compute \textsc{BSS Eval} scores on the different mixtures (for synthetic and real data) and on 30 different initializations (for MIDI data).

The numbers of iterations are chosen such that the performance is not further improved beyond. Energy ratios are expressed in dB.

\section{Experimental results}
\label{sec:exp}

\subsection{HRNMF initialization and estimation algorithm}
\label{sec:HRNMF_init}
HRNMF requires a well-chosen initialization to produce meaningful results (likely because of the great number of local minima of the cost function). The data to be processed is a mixture of real notes without frequency overlap. We consider the regular EM algorithm \cite{Badeau2011} and the VBEM algorithm \cite{Badeau2014}. Initializations can be random, KLNMF \cite{Lee2001} or Itakura-Saito NMF (ISNMF, \cite{Fevotte2009}), computed by means of MUR algorithms.

\begin{table}[ht]
\center
\caption{Influence of HRNMF initialization and algorithm on source separation performance}
\label{tab:HRNMF_initialization}
\begin{tabular}{c|c|c|c|c|c}

Algorithm & Initialization & SDR & SIR & SAR & Time (s) \\
\hline
\multirow{3}{*}{EM}
& Random   & $5.3$  & $6.4$ & $14.3$ & $379$\\
& ISNMF    & $15.0$  & $21.2$ & $17.0$ & $376$ \\
& KLNMF    & $17.0$  & $22.2$ & $18.7$ & $377$ \\
 \hline
\multirow{3}{*}{VBEM} 
& Random & $1.4$  & $2.8$ & $11.1$ & $1.03$ \\
& ISNMF  & $\mathbf{16.9}$ & $\mathbf{25.3}$ & $\mathbf{17.7}$ & $\mathbf{0.95}$\\
& KLNMF  & $\mathbf{16.9}$ & $\mathbf{24.5}$ & $\mathbf{17.8}$ & $\mathbf{0.89}$\\
 \hline

\end{tabular}
\end{table}

Results are presented in Table~\ref{tab:HRNMF_initialization} (the best performance is highlighted in bold font).
We observe that initializing HRNMF with a prior NMF algorithm provides significantly better results than applying the EM or VBEM algorithm directly on random parameters. The choice of the NMF (KL or IS) does not influence much the results. We also see that the VBEM algorithm provides results similar to the EM algorithm, with a dramatic reduction of the computational cost. We will thus use the VBEM algorithm with KLNMF initialization for our benchmark.

\subsection{Synthetic data}

\begin{figure}[ht]
\centering
\includegraphics[scale=0.45]{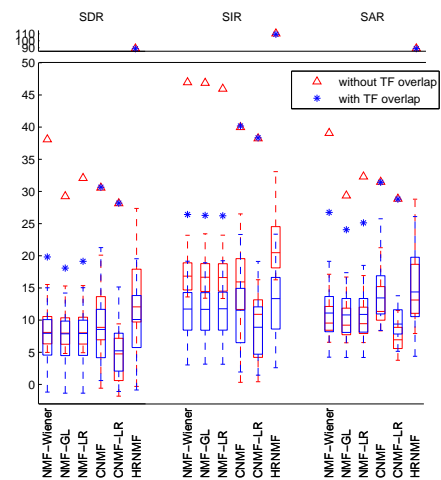}
\caption{Synthetic harmonics separation performance}
\label{fig:synth_benchmark}
\end{figure}

Benchmark results for synthetic harmonics are presented in Figure \ref{fig:synth_benchmark}. Box-plots compile data for blind approach. Each box-plot is made up of a central line indicating the median of the data, upper and lower box edges indicating the $1^{st}$ and $3^{rd}$ quartiles, and whiskers indicating the minimum and maximum values. The triangles and stars indicate the performance of the oracle approach.

These results show that \textbf{Griffin-Lim} and \textbf{LeRoux} phase reconstruction algorithms provide poor results in terms of audio quality. While consistency is increased in \textbf{NMF-GL} and \textbf{NMF-LR}, those methods lead to a decrease of the SDR and SAR scores compared to \textbf{NMF-Wiener}. Enforcing the magnitude to be constant over iterations seems too constraining to increase audio quality. \textbf{CNMF-LR} is supposed to be a response to the aforementioned problem, but it does not provide better results than \textbf{NMF-LR}. It also requires much more memory for storing the phase field of each source. We also note that \textbf{CNMF} provides better results than \textbf{CNMF-LR}, confirming that consistency may not be a good criterion for audio quality. Results generally drop when TF bins overlap, but not in terms of SAR: artifact rejection seems globally increased when overlap occurs in the blind benchmark.

Finally, blind separation with the \textbf{HRNMF} model provides slightly better results than the other models (except when overlap occurs in the TF domain: \textbf{CNMF} and \textbf{HRNMF} then lead to a similar SAR median). This model also provides the best performance in the oracle benchmark. \textbf{NMF-Wiener} is the fastest algorithm ($40$ ms), the other models are estimated in approximately $1.5$ seconds. Similar computation times are obtained for real data.

The tests performed on synthetic harmonics with vibratos (that cannot be presented here because of a lack of room) lead to similar results: the \textbf{HRNMF} model significantly outperforms the other models in the oracle approach, demonstrating its ability to accurately represent a variety of signals.

\subsection{Piano notes mixtures}
\label{sec:real_data}

\begin{figure}[ht]
\centering
\includegraphics[scale=0.5]{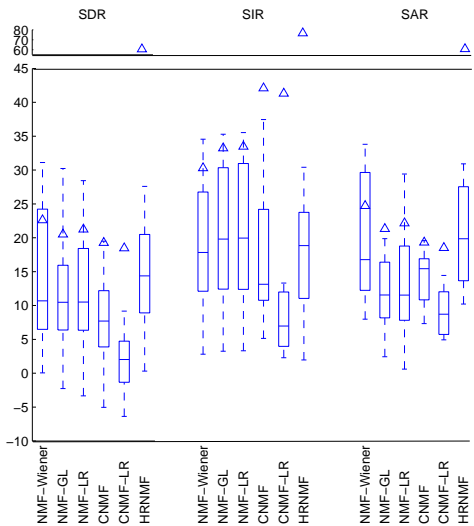}
\caption{Piano notes mixtures separation performance}
\label{fig:real_benchmark}
\end{figure}

Benchmark results for piano notes mixtures are presented in Figure \ref{fig:real_benchmark}.
We note that the algorithms do not perform worse than in the synthetic data case. 
The blind benchmark shows that \textbf{HRNMF} results are similar to the other algorithms (or slightly better), but the oracle results confirm that it is the best model available in terms of potential for source separation.
\textbf{NMF-Wiener} is also interesting, because it provides a fast and relatively accurate audio source separation. The analysis of the results for each mixture reveals that the quality of \textbf{NMF-Wiener} is slightly worse than \textbf{HRNMF} when there are overlapping TF bins.

\subsection{MIDI song}


\begin{figure}[ht]
\centering
\includegraphics[scale=0.5]{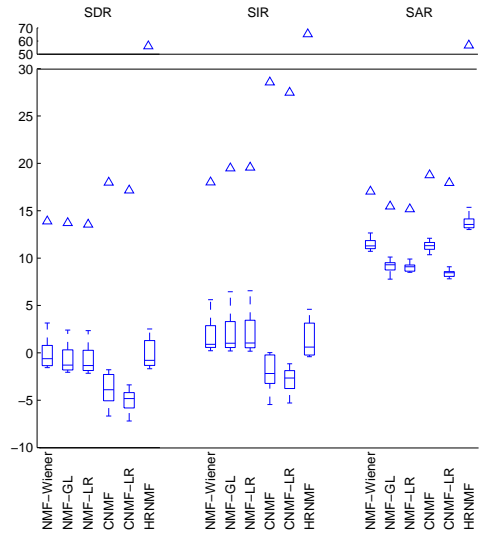}
\caption{MIDI audio excerpt separation performance}
\label{fig:midi_benchmark}
\end{figure}

Figure~\ref{fig:midi_benchmark} presents the results obtained with a realistic MIDI audio excerpt. It shows a dramatic reduction of blind source separation quality compared to the previous tests. This signal seems too complex to obtain an efficient factorization. Then, \textbf{HRNMF} estimation does not improve the result of the initial KLNMF. However, the oracle approach still shows that this method has a higher potential than the other models for this application. \textbf{NMF-Wiener} is computed in $60$ ms and the others models are estimated in $3$ to $4$ seconds.

\section{Conclusion}
\label{sec:conclu}

This benchmark presents HRNMF as a very promising model in terms of source separation quality. It is able to take both phases and correlations over time into account, and models a variety of signals frequently observed in music analysis. In particular, the oracle approach showed that HRNMF is likely to be particularly efficient when source separation is partially informed. The other models and algorithms appear to be less appealing for source separation, because sources often overlap in the TF domain, a common phenomenon in music. More generally, algorithms that take correlation over time and frequencies into account with a great expressive power should be considered with particular attention. Consistency has also been shown not to be an appropriate criterion for audio quality. The datasets and procedure described in this work can be a good basis for further evaluation of the potential of source separation models.

Besides, the experiments show that the VBEM algorithm used for estimating HRNMF is highly sensitive to initialization. Semi-supervised learning or prior information about the sources, such as harmonicity, sparsity or temporal smoothness should be introduced in order to address this issue. Alternative estimation methods, more robust and less sensitive to initialization, could be implemented in future research. Bayesian methods such as Markov Chain Monte Carlo (MCMC) and message passing algorithms might be an option. Alternatively, the algebraic principles used in High Resolution methods (such as the ESPRIT algorithm \cite{Hua2004}) could also be exploited in order to address this estimation problem.

\newpage


\bibliographystyle{IEEEbib}
\bibliography{references_icassp2015}

\end{document}